\documentclass[prb,twocolumn,superscriptaddress,showpacs]{revtex4-1}
\usepackage{graphics}
\usepackage{subfigure}
\usepackage{graphicx}   
\usepackage{amsmath}
\usepackage{amstext}
\usepackage{amssymb}
\usepackage{color}
\usepackage{pstricks}
\newcommand{\mbf}[1]{\mathbf{#1}} 
\newcommand{\kk}{\mbf{k}}

\newcommand{\q}{\mbf{q}}
\newcommand{\CommentOut}[1]{}

\newcommand{\ut}[1]{{\rm\; #1}}
\graphicspath{{./figs/pdffigs/}}

\begin{document}

\title{Superconducting gap in LiFeAs from three-dimensional spin-fluctuation pairing calculations}

\author{Y. Wang}
\author{A. Kreisel}
\affiliation{Department of Physics, University of Florida, Gainesville, Florida 32611, U.S.A.}

\author{V. B. Zabolotnyy}
\affiliation{Leibniz-Institute for Solid State Research, IFW-Dresden, D-01171 Dresden, Germany}
\affiliation{Physikalisches Institut, EP IV, Universit\"{a}t W\"{u}rzburg, D-97074 W\"{u}rzburg, Germany}

\author{S. V. Borisenko}
\affiliation{Leibniz-Institute for Solid State Research, IFW-Dresden, D-01171 Dresden, Germany}

\author{B.~B\"{u}chner}
\affiliation{Leibniz-Institute for Solid State Research, IFW-Dresden, D-01171 Dresden, Germany}
\affiliation{Institut f\"{u}r Festk\"{o}rperphysik, Technische Universit\"{a}t Dresden, D-01171 Dresden,
Germany}

\author{T. A. Maier}
\affiliation{Center for Nanophase Materials Sciences and Computer Science and Mathematics Division, Oak Ridge
National Laboratory, Oak Ridge, Tennessee 37831-6494}

\author{P. J. Hirschfeld}
\affiliation{Department of Physics, University of Florida, Gainesville, Florida 32611, U.S.A.}

\author{D. J. Scalapino}
\affiliation{Department of Physics, University of California, Santa Barbara, California 93106-9530 USA}

\date{\today}

\begin{abstract}
The lack of nesting of the electron and hole Fermi-surface sheets in the Fe-based superconductor LiFeAs, with
a critical temperature of 18~K, has led to questions as to whether the origin of superconductivity in this
material might be different from other Fe-based superconductors. Both angle-resolved photoemission and
quasiparticle interference experiments have reported fully gapped superconducting order parameters with
significant anisotropy. The system is also of interest because relatively strong correlations seem to be
responsible for significant renormalization of  the hole bands. Here we present calculations of the
superconducting gap and pairing  in the random-phase approximation using Fermi surfaces derived from measured
photoemission spectra. The qualitative features of the gaps obtained in these calculations are shown to be
different from previous two-dimensional theoretical works and in good agreement with experiment on the main
Fermi surface pockets. We analyze the contributions to the pairing vertex thus obtained and show that the
scattering processes between electron and hole pockets that are believed to  dominate the pairing in other
Fe-based superconductors continue to do so in LiFeAs despite the lack of nesting, leading to gaps with
anisotropic $s_\pm$ structure. Some interesting differences relating to the enhanced $d_{xy}$ orbital content
of the LiFeAs Fermi surface are noted.
\end{abstract}

\maketitle

\section{Introduction} \label{sec:1}
\begin{figure}
\includegraphics[width=1\columnwidth]{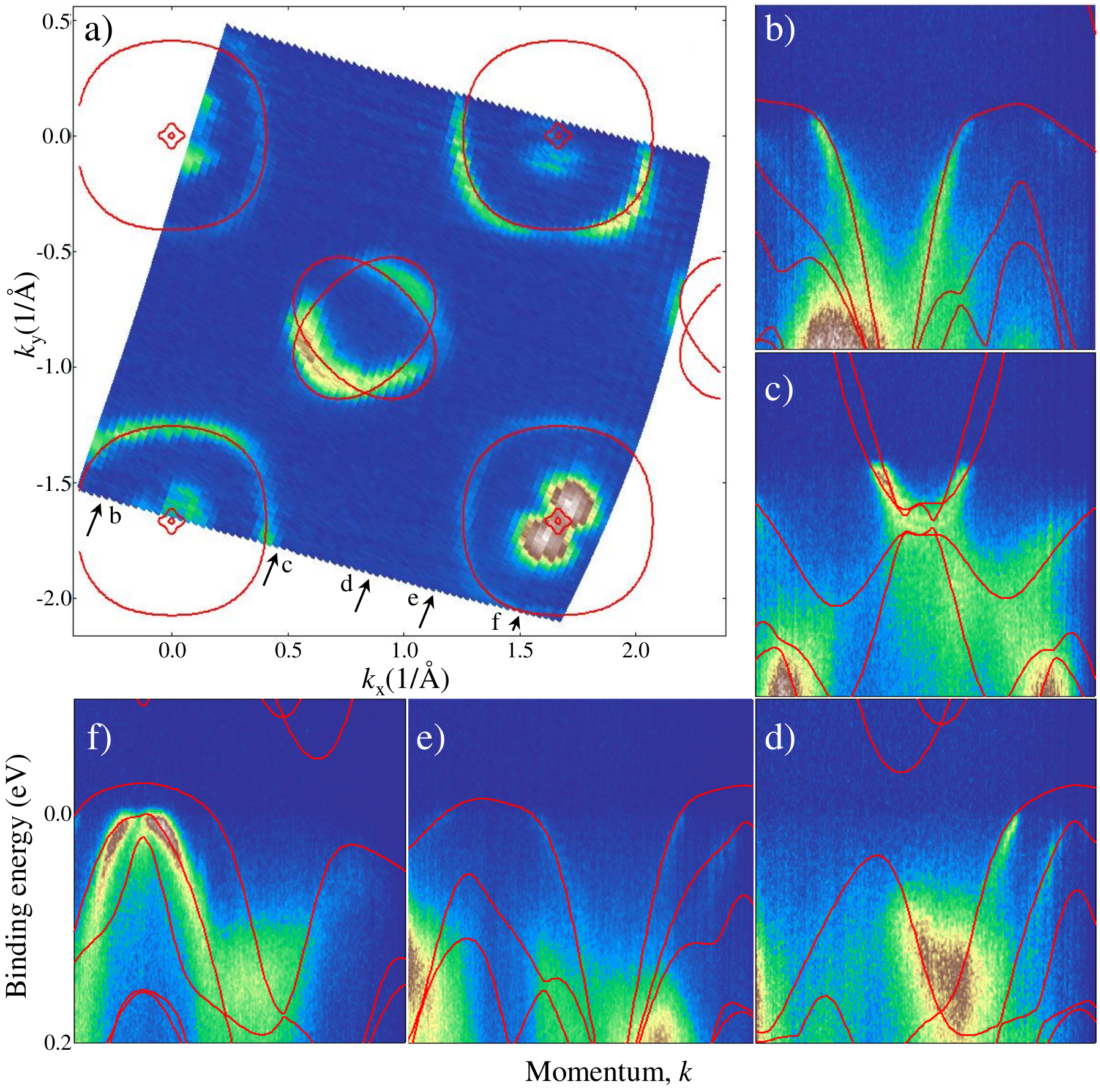}
\caption{(Color online) Comparison of the tight-binding bands and ARPES data both (a) at the Fermi surface
and (b)--(f) in energy--momentum cuts for $k_z = \pi/2$. The black arrows in (a) denote the positions of
several representative energy momentum cuts. For demonstration purposes here we use one of our high-quality
Fermi-surface maps from Ref.~\onlinecite{borisenko_Symmetry}, although to recover additional information on
$k_z$ dispersion more data with various $h\nu$ were used. For further details see
Appendix~\ref{appendix:TBmodel}.}
 \label{fig:fit}
\end{figure}

\begin{figure}
  \includegraphics[width=0.9\columnwidth]{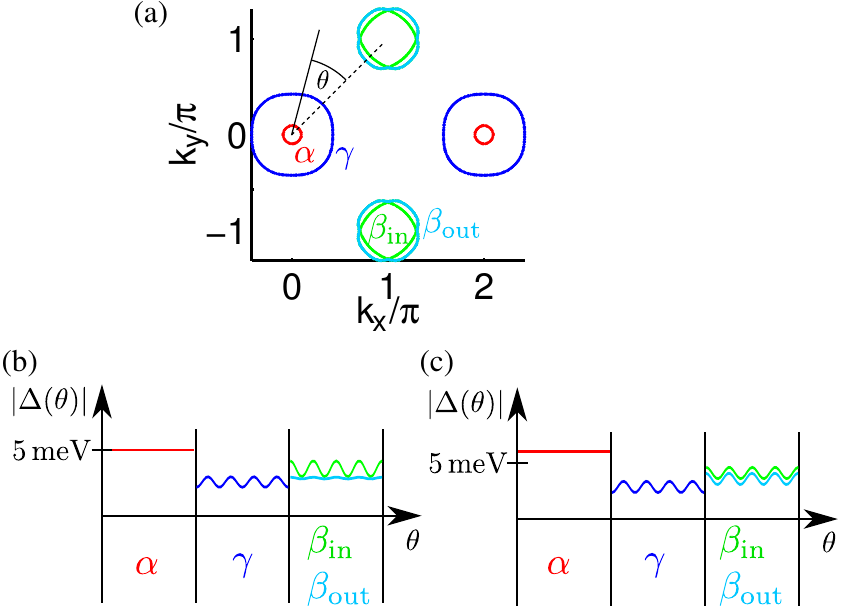}
\caption{(Color online) (a) The cut of the Fermi surface of the ARPES-derived tight-binding model (filling
$n=6.00$) at $k_z=\pi$ to show the definition of the various pockets and the angle $\theta$ that parametrizes
the surface points. Sketch of the results of the gap $|\Delta(\theta)|$ as seen in recent ARPES experiments
compiled from the fits provided in (b) Ref.~\onlinecite{Umezawa_2012} and (c)
Ref.~\onlinecite{borisenko_Symmetry}.}
 \label{fig:cartoon}
\end{figure}

The compound LiFeAs is an 18~K superconductor that presents several novel features relative to the other
families of Fe pnictides~\cite{StewartReview}. High-quality crystals with atomically flat nonpolar surfaces
are now straightforward to prepare, and the surface electronic structure has been shown to be the same as in
the bulk~\cite{HEschrigLiFeAs}, suggesting that this system and related 111 materials are ideal ones to apply
surface spectroscopies like angle-resolved photoemission (ARPES) and scanning tunneling microscopy
(STM)~\cite{STMLiFeAs}. ARPES experiments~\cite{borisenko_2010,Umezawa_2012,Hajiri12}  and electronic
structure calculations within density functional theory (DFT)~\cite{HEschrigLiFeAs,Hajiri12,Nekrasov08}
reported early on a Fermi surface very different from the  conventional set of hole and electron pockets
predicted by DFT for the other Fe-based superconductors (Figs.~\ref{fig:fit} and \ref{fig:cartoon}).  In
particular, less clear nesting of hole and electron pockets was observed, leading to the suggestion that this
was the reason for the absence of magnetism in this parent compound~\cite{borisenko_2010}.  More recently, de
Haas-van Alphen (dHvA) measurements~\cite{Putzke} showed reasonable agreement with bulk DFT for orbits on the
electron pockets.

One continuing puzzle has been the small to negligible size of the inner ($\alpha_1$, $\alpha_2$) hole
pockets observed by ARPES compared to the relatively large sizes found in DFT. Recently, local-density
approximation (LDA) + dynamical mean-field theory (DMFT) calculations have presented a picture which suggests
that the 111 are considerably more correlated than, e.g., the well-studied 122 materials and have argued that
stronger interactions lead to a shrinkage of the inner hole pockets but maintenance of the electron pocket
size and shape~\cite{KotliarNatMat,Ferberetal,Haule12}. This picture would then account for both ARPES and
dHvA results, including very recent dHvA measurements which detected very small holelike
orbits~\cite{bZeng13}. However, the extent of the agreement of LDA + DMFT theory and experiment for the hole
pockets is obscured somewhat by disagreements among the various calculations as to the size of the inner
pockets, as well as by the challenges of resolving the near-grazing $\Gamma$-centered hole bands in ARPES.

Within the spin-fluctuation model for pairing in the Fe-based materials, the structure of the Fermi surface
is crucial for superconductivity as well as magnetism.  Since the usual arguments leading to $s_\pm$
pairing~\cite{Mazin_spm} invoke interband pair scattering between electron and hole pockets enhanced by
nesting, the absence of nesting in this material would seem to undercut the case for an $s_\pm$
superconducting state. A second aspect of this discussion  relates to the spin symmetry of the order
parameter.  While early NMR work reported a strongly temperature-dependent Knight shift and $1/T_1$ below
$T_c$, consistent with $s$-wave pairing~\cite{Li_etal_LiFeAs_NMR},  Baek \textit{et
al.}~\cite{Buechner_LiFeAsNMR} reported a Knight shift in some magnetic field directions with no $T$
dependence, suggestive of equal spin-triplet pairing, which would then be consistent  with  theoretical
analysis proposing triplet pairing for this system~\cite{vdBrink_triplet}.  Neutron experiments have thus far
not provided conclusive evidence one way or another.  A weak incommensurate spin resonance was observed in
inelastic neutron scattering experiments~\cite{Taylor_neutron_LiFeAs} and associated with a probable $s_\pm$
state, but it should be noted that the existence of a spin resonance does not definitively exclude triplet
pairing~\cite{Morr}.

More recently, some authors~\cite{borisenko_Symmetry} reported detailed ARPES measurements of the
superconducting gap in LiFeAs. These measurements were remarkable in the sense that while they showed that
the system has a full gap, consistent with other low-temperature probes~\cite{LiFeAspendepth, borisenko_2010,
tanatar_2011, STMLiFeAs, Shibauchi_LiFeP_2011}, they also exhibited substantial gap anisotropy around both
electron and hole Fermi surfaces.  Similar but not quantitatively identical results for anisotropic gaps were
reported by Umezawa \textit{et al.}~\cite{Umezawa_2012}  The reports of strongly angle-dependent gaps are
relatively rare among the many ARPES measurements on Fe-based superconductors (for exceptions see
Refs.~\onlinecite{ShinKFe2As2} and~\onlinecite{Shin12}), where isotropic gaps are often reported even for
those systems where it is believed from low-temperature transport measurements that gap nodes exist (for a
discussion of this so-called ``ARPES paradox,'' see Ref.~\onlinecite{HKM_review}). The existence of
anisotropy around some of the Fermi-surface pockets of LiFeAs was also reported by Allan \textit{et
al.}~\cite{AllenetalScience}, who performed high-resolution, low-temperature STM measurements together with a
quasiparticle interference (QPI) analysis which found a small gap nearly identical to ARPES on the large
outer hole ($\gamma$)  pocket, with gap minima along the Fe-Fe bond direction (as suggested in
Ref.~\onlinecite{YanWangvortex}).  A second, larger gap, also with moderate anisotropy, was reported and
attributed to an inner $\alpha_{1,2}$ hole pocket.

To illustrate the types of gaps found by the ARPES experiments,  we present in Fig. \ref{fig:cartoon} a
schematic representation of these data to familiarize the reader with the qualitative features reported.  One
can see that several aspects stand out: (a) oscillatory gaps on the outer hole ($\gamma$) and electron
($\beta$) pockets, (b) smallest gap on the $\gamma$ pocket, and (c) large gaps of roughly equal average size
on inner hole ($\alpha$) and electron  ($\beta$) pockets.   The relative phases of the gap oscillations on
the two $\beta$ pockets are also striking. We note here that the measurement of the gap on the $\alpha$
pocket is particularly delicate since this band barely crosses the Fermi level near the $Z$ point, and may
not cross near $\Gamma$ at all.

It is essential to the understanding of superconductivity in Fe-based superconductors to decide whether
LiFeAs fits into the usual framework, with pairing driven by spin fluctuations, or represents different
physics.  Testing to see if the various qualitative and quantitative features of the gaps reported in
experiment can be reproduced is therefore an important challenge to theory.  In this paper we calculate the
effective pairing vertex within the fluctuation exchange approximation for the full three-dimensional (3D)
Fermi surface of LiFeAs and compare our results for the superconducting states which become stable at the
transition to previous theory and to experiment.  To understand how robust these results are, we perform the
calculation for a band structure fit to the ARPES results, which differ  primarily from DFT due to the much
larger size of the inner hole pocket in the latter, as discussed above, as well as large shifts in the
orbital character of the Fermi surfaces.  In addition, we compare our results to a slightly hole-doped system
to simulate the effect of missing Li at the LiFeAs surface and to calculations with a ``standard"  DFT band
structure. We find that most aspects of the superconducting gap are remarkably well reproduced by the theory
using the ARPES-derived electronic structure model.   Our conclusion is that the superconductivity in LiFeAs
is very likely to be of the ``conventional'' $s_\pm$ type, with significant anisotropy on both hole and
electron pockets.

\section{Ten-orbital tight-binding fits and Fermi surfaces}
\begin{figure*}
 \centering
 \includegraphics[width=0.95\textwidth,draft=false]{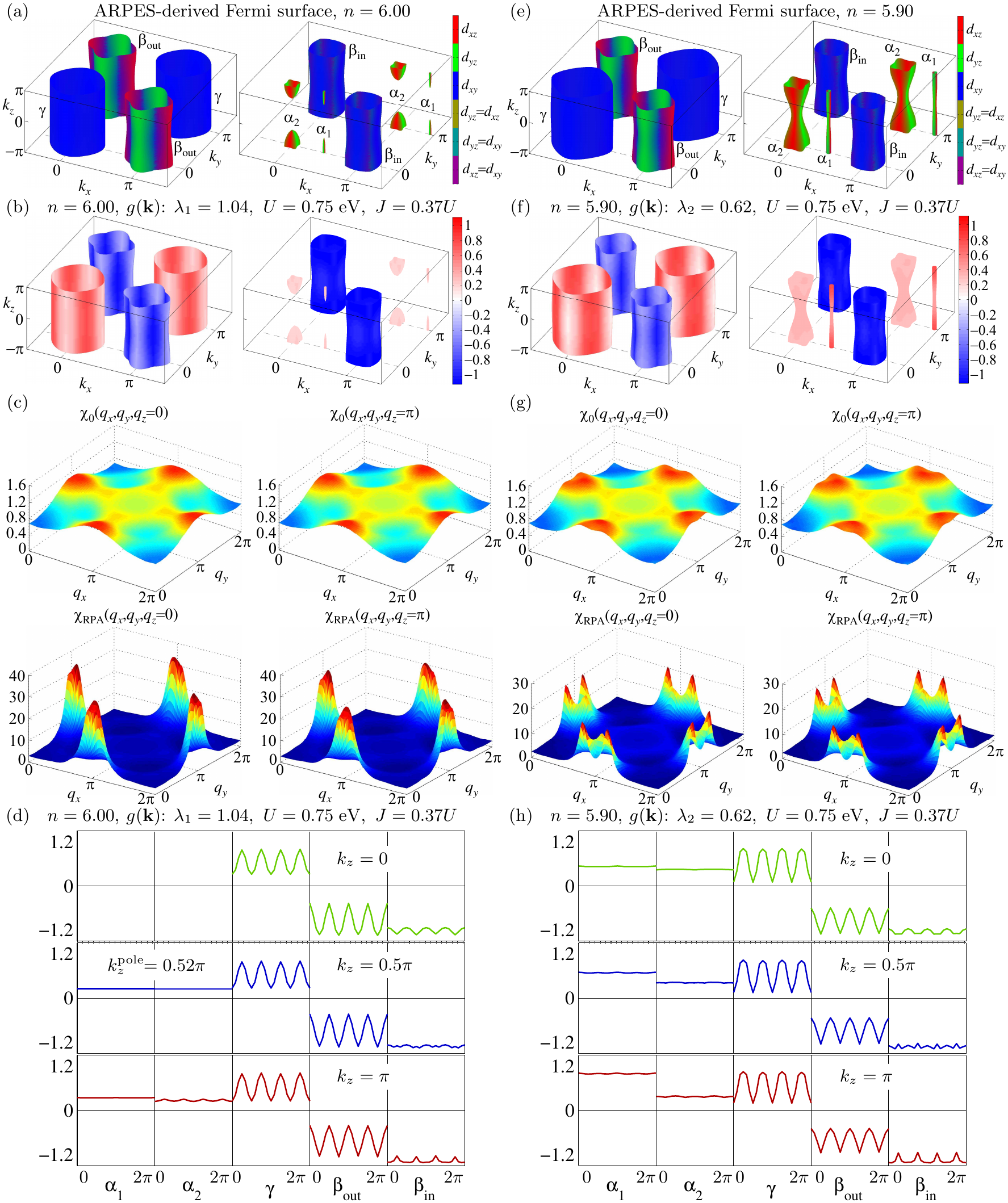}
\caption{(Color online) Fermi surface of LiFeAs from ten-orbital $H_0^\text{ARPES}$ at (a) filling $n=6.00$
and (e) $n=5.90$ plotted in the coordinates of the one-Fe Brillouin zone as two sets, outer (left) and inner
(right) pockets. Majority orbital weights are labeled by colors, as shown. Note the small innermost, hole
pocket $\alpha_1$ with rotation axis $\Gamma$-$Z$ ($M$-$A$) has been artificially displaced from its position
along the $k_x$ axis for better viewing in (a) and (e). (b) and (f) are the gap symmetry functions
$g(\mathbf{k})$ corresponding to the leading eigenvalues ($s_\pm$ wave) and interaction parameters shown in
the figure. (c) and (g) are the corresponding noninteracting spin susceptibility
[$\chi_s(\mathbf{q},\omega=0)$ for $U=0,\; J=0$] and RPA spin susceptibility [$\chi_s(\mathbf{q},\omega=0)$
for the same $U,\; J$ as in (b) and (f)] at $q_z=0,\pi$. In the RPA susceptibility plot, a thin white line is
plotted along the path $(\pi,q_y,q_z=0)$ or $(\pi,q_y,q_z=\pi)$, its projection on the $q_y$-$\chi_s$ plane
is plotted as a thick orange line, and the red triangle indicates the peak position. (d) and (h) are the
angle dependence of $g(\mathbf{k})$ on the pockets indicated at $k_z=0,0.5\pi,\pi$. In (d) the gap value on
$\alpha$ pockets at the pole is plotted since these pockets do not extend to $k_z=0.5\pi$.}
 \label{fig:FSGapsSusc}
\end{figure*}

Our approach here to the pairing calculation differs somewhat from those performed for materials where DFT
and ARPES were in qualitatively good agreement.  Since the spin-fluctuation pairing theory involves states
very close to the Fermi surface, the disagreement between DFT and ARPES suggests that strong electronic
correlations must be accounted for at some level.   The simplest modification of the usual approach is to
adopt a band structure which fits experiment well, a procedure which is not uniquely defined due to the
multiband nature of the system. We have chosen to begin with a ten-Fe orbital tight-binding Hamiltonian
$H_0^\text{ARPES}$,  fit to measured ARPES data on a high-quality LiFeAs sample~\cite{borisenko_Symmetry}
using the method of Ref.~\onlinecite{Eschrig09}, which we refer to as the ARPES-derived band structure. The
hopping parameters and the dispersions are given in the Appendix C, and the comparison of the tight binding
bands and Fermi surface cuts are  shown in Fig.~\ref{fig:fit}.  The full Fermi surface from this model is
then shown in Figs.~\ref{fig:FSGapsSusc}(a) and  \ref{fig:FSGapsSusc}(e) for two different dopings, $n=6.00$
and $n=5.90$. The latter results are presented to mimic the possible effects of Li deficiency which are known
to be present in the sample and because the Fermi-surface topology changes abruptly near $n=6.00$.  We find
that these changes are potentially quite important for the superconductivity, as discussed below.

It is interesting to first compare the ARPES-derived  Fermi surface in Fig.~\ref{fig:FSGapsSusc}(a)  to the
DFT Fermi surface  discussed in the Appendix A since the DFT results are essentially those used in earlier
two-dimensional (2D) spin-fluctuation calculations~\cite{Platt11}.  Both the DFT- and the ARPES-derived Fermi
surfaces include similar large hole pockets ($\gamma$) and inner and outer electron pockets
($\beta_\text{in}$, $\beta_\text{out}$). The $\gamma$ pockets are of comparable size and are similar in
shape. In the DFT-derived model, the inner and outer $\beta$ pockets cross each other along high symmetry
directions equivalent to $X$-$Y$ in the one-Fe zone. They also approach each other closely at nonzero $k_z$
values away from the high-symmetry directions due to the hybridization of the DFT electron bands, and this
leads to some $k_z$ distortions and abrupt changes in their orbital characters with $k_z$. By contrast, the
$k_z$-dispersions of the ARPES-derived electron pockets are weak. The pockets only approach each other at the
high symmetry directions (where they cross in the absence of spin-orbit coupling), and they retain their
orbital characters along $k_z$, as measured by the ARPES experiment.~\cite{borisenko2013} The main difference
beyond these shifts of orbital characters and shape of the outer $\beta$ Fermi sheets is the much smaller
$\alpha_{1,2}$ hole pockets and the closing of the $\alpha_2$ hole pocket in the ARPES-derived inner-hole
Fermi sheets. The density of states (DOS) at the Fermi level is shown in Table~\ref{tab:dosDFTandARPES} in
Appendix A. Within a scaling factor $r=0.5$, the total densities of states and partial density of states are
quite consistent between these two models.

The calculated carrier concentration in the compensated ($n=6.00$) case (number of electrons/Fe = number of
holes/Fe) from $H_0^\text{ARPES}$ is roughly consistent with 0.18 electrons/Fe and 0.2 holes/Fe from the
ARPES experiment by Umezawa \textit{et al.}~\cite{Umezawa_2012}.  It is interesting to note that the
difference in hole and electron carriers in Ref.~\onlinecite{Umezawa_2012} is already a hint that the surface
of the sample may contain some Li vacancies and therefore be slightly hole doped. For the $n=5.90$ case we
have chosen here for illustration's sake corresponding electron and hole densities that are 0.16 and 0.26,
respectively.

\begin{figure}
 \centering
 \includegraphics[width=1\columnwidth,draft=false]{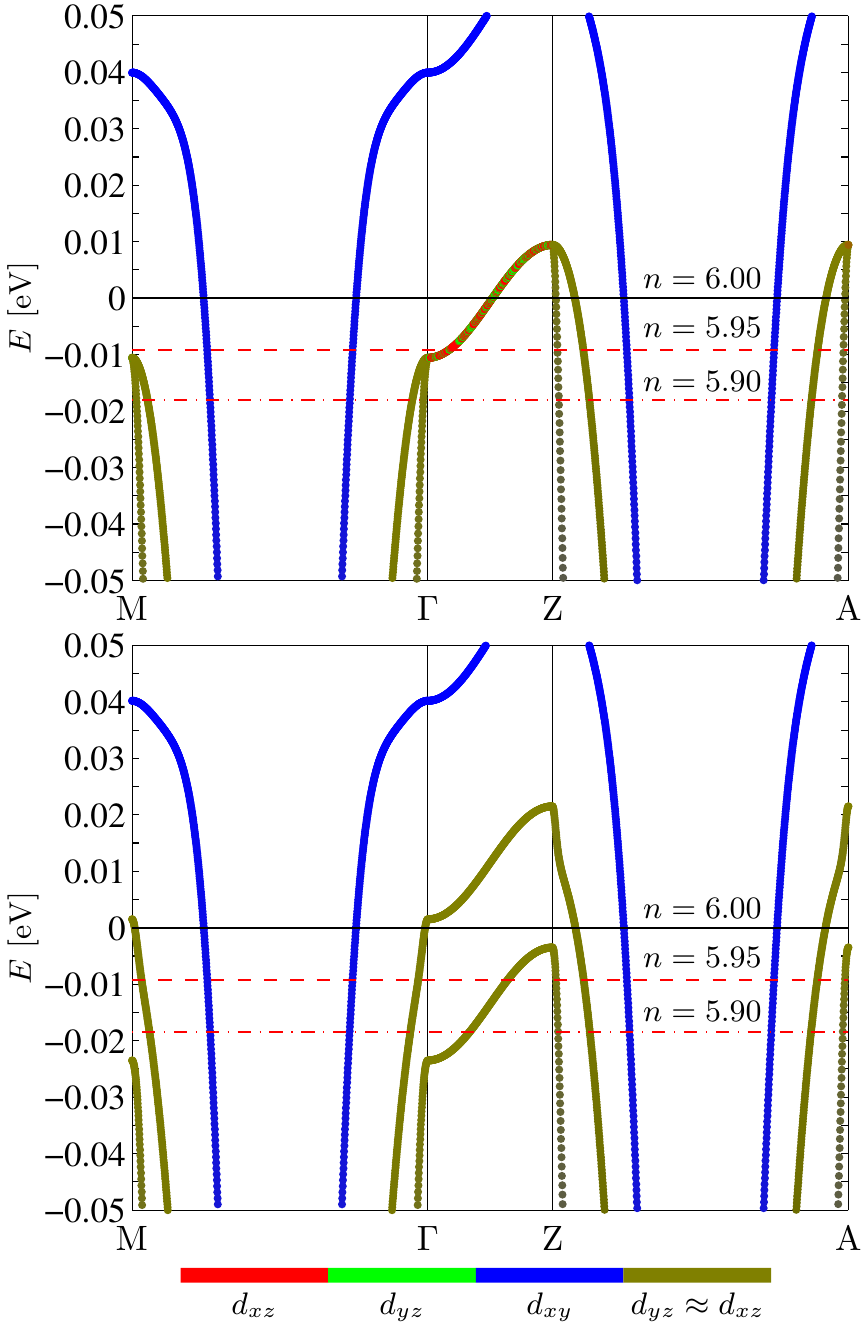}
\caption{(Color online) The band structures along the $M$-$\Gamma$-$Z$-$A$ path in the one-Fe Brillouin zone
for (top) the ARPES-derived model and (bottom) the same model with the approximate spin-orbit coupling
term~\cite{Kreisel13} $\lambda_\text{Fe}^{3d}\sum_{i}L_i^zS_i^z$, with $\lambda_\text{Fe}^{3d}=0.025$~eV. The
color encodes the major orbital characters, as indicated by the horizontal color bar. The dashed lines mark
the corresponding Fermi energy at filling $n=6.00$, $5.95$, and $5.90$.}
 \label{fig:Bandsdoping}
\end{figure}

In general, the ARPES-derived tight-binding model is a close fit to the ARPES data in
Ref.~\onlinecite{borisenko_Symmetry} and reproduces the orbital characters on all pockets.  One apparently
minor discrepancy (which may play a more important role than expected at first sight; see below) is that due
to the crystal symmetry, the two hole bands dispersing near $Z$  in the tight-binding Hamiltonian
$H_0^\text{ARPES}$ are degenerate at the $Z$ point and therefore in a nonrelativistic calculation must both
cross or neither cross the Fermi surface, as shown in Fig.~\ref{fig:Bandsdoping} (top panel). Apart from the
large $\gamma$ pocket, ARPES observes only a single holelike band ($\alpha_2$) crossing the Fermi surface
near $Z$, while a second holelike band ($\alpha_1$) is pushed below. This suggests that spin-orbit coupling,
which will split the two hole bands as shown in Fig.~\ref{fig:Bandsdoping} (bottom panel), may be relevant
here. For the moment we neglect this distinction and focus on the nonrelativistic band structure, but we will
return to it in the discussion below.

\section{Fluctuation exchange pairing model}
With the tight-binding Hamiltonian $H_0$ in the previous section, we include the local interaction via the
ten-orbital Hubbard-Hund Hamiltonian
\begin{align}
  H = H_0 &+ \bar{U}\sum_{i,\ell}n_{i\ell\uparrow}n_{i\ell\downarrow}
           + \bar{U}'\sum_{i,\ell'<\ell}n_{i\ell}n_{i\ell'}     \notag\\
  &+\bar{J}\sum_{i,\ell'<\ell}\sum_{\sigma,\sigma'}
           c_{i\ell\sigma}^\dag c_{i\ell'\sigma'}^\dag c_{i\ell\sigma'} c_{i\ell'\sigma}     \notag\\
  &+\bar{J}'\sum_{i,\ell'\neq\ell}
           c_{i\ell\uparrow}^\dag c_{i\ell\downarrow}^\dag c_{i\ell'\downarrow} c_{i\ell'\uparrow}, \label{eq:HubbardH}
\end{align}
where the interaction parameters $\bar{U}$, $\bar{U}'$, $\bar{J}$, $\bar{J}'$ are given in the notation of
Kuroki \textit{et al.}~\cite{Kuroki08} Here $\ell$ is the orbital index corresponding to Fe $3d$-orbitals,
and $i$ is the Fe \emph{atom} site. The spectral representation of the one-particle Green's function is given
as
\begin{align}
  G_{\ell_1\ell_2}(\mathbf{k},i\omega_n) &= \sum_\mu
    \frac{a_\mu^{\ell_1}(\mathbf{k})a_\mu^{\ell_2,*}(\mathbf{k})}{i\omega_n-E_\mu(\mathbf{k})}, \label{eq:GreenFunction}
\end{align}
where the matrix elements $a_\mu^{\ell}(\mathbf{k})=\langle \ell|\mu\mathbf{k} \rangle$ are spectral weights
of the Bloch state $|\mu\mathbf{k} \rangle$ with band index $\mu$ and wave vector $\mathbf{k}$ in the orbital
basis and $\omega_n$ are the fermionic Matsubara frequencies. In terms of the Green's function, the orbitally
resolved noninteracting susceptibility is
\begin{align}
  &\chi^0_{\ell_1\ell_2\ell_3\ell_4}(\mathbf{q},i\omega_m) \notag\\
  &\quad = -\frac{1}{N\beta}\sum_{\mathbf{k},i\omega_n}
    G_{\ell_4\ell_2}(\mathbf{k},i\omega_n) G_{\ell_1\ell_3}(\mathbf{k+q},i\omega_n+i\omega_m), \label{eq:chi0a}
\end{align}
where $N$ is the number of Fe atom sites, $\beta=1/T$ is the inverse temperature and $\omega_m$ are the
bosonic Matsubara frequencies. After summing the fermionic Matsubara frequencies following the analytic
continuation to the real axis of bosonic Matsubara frequencies, we obtain the retarded susceptibility
\begin{align}
  &\chi^0_{\ell_1\ell_2\ell_3\ell_4}(\mathbf{q},\omega) \notag\\
  &\quad = -\frac{1}{N}\sum_{\mathbf{k},\mu\nu}
    \frac{a_\mu^{\ell_4}(\mathbf{k})a_\mu^{\ell_2,*}(\mathbf{k})a_\nu^{\ell_1}(\mathbf{k+q})a_\nu^{\ell_3,*}(\mathbf{k+q})}
    {\omega+E_\mu(\mathbf{k})-E_\nu(\mathbf{k+q})+i0^+} \notag\\
  &\quad\quad \times\left\{ f[E_\mu(\mathbf{k})] - f[E_\nu(\mathbf{k+q})] \right\}. \label{eq:chi0b}
\end{align}
For the 3D $\mathbf{k}$-sum we use a ($47\times 47\times 31$)-point $\mathbf{k}$ mesh for the ARPES-derived
model in the one-Fe Brillouin zone (1Fe-BZ); we interpolate the static noninteracting susceptibility
$\chi^0_{\ell_1\ell_2\ell_3\ell_4}(\mathbf{q},\omega=0)$ from directly calculated susceptibility values on a
($20\times 20\times 8$)-point $\mathbf{q}$ mesh in the 1Fe-BZ to perform the expensive numerical calculation
with a fine patched Fermi surface in solving the pairing eigenvalue problem. Within the random-phase
approximation (RPA) we define the spin-fluctuation ($\chi_1^\text{RPA}$) and orbital-fluctuation
($\chi_0^\text{RPA}$) parts of the RPA susceptibility as
\begin{subequations}
\begin{align}
  \chi_{1,\ell_1\ell_2\ell_3\ell_4}^\text{RPA}(\mathbf{q},\omega) &=
    \left\{ \chi^0(\mathbf{q},\omega)[1-\bar{U}^s\chi^0(\mathbf{q},\omega)]^{-1}
    \right\}_{\ell_1\ell_2\ell_3\ell_4}, \label{eq:chiRPAs} \\
  \chi_{0,\ell_1\ell_2\ell_3\ell_4}^\text{RPA}(\mathbf{q},\omega) &=
    \left\{ \chi^0(\mathbf{q},\omega)[1+\bar{U}^c\chi^0(\mathbf{q},\omega)]^{-1}
    \right\}_{\ell_1\ell_2\ell_3\ell_4},
\end{align}
\end{subequations}
such that the RPA-enhanced spin susceptibility is then given by the sum
\begin{align}
  \chi_s(\mathbf{q},\omega) = \frac{1}{2}\sum_{\ell_1 \ell_2}\chi_{1,\ell_1\ell_1\ell_2\ell_2}^\text{RPA}(\mathbf{q},\omega).
\end{align}
The interaction matrices $\bar{U}^s$ and $\bar{U}^c$ in orbital space have matrix elements consisting of
linear combinations of the interaction parameters, and their explicit forms are given, e.g., in
Ref.~\onlinecite{Kemper10}.

Next, we define the singlet pairing vertex in band space,
\begin{align}
  &\Gamma_{ij}(\mathbf{k,k'}) = \mathrm{Re}\sum_{\ell_1\ell_2\ell_3\ell_4}
    a_{\nu_i}^{\ell_1,*}(\mathbf{k})a_{\nu_i}^{\ell_4,*}(\mathbf{-k}) \notag \\
  &\quad\times [\Gamma_{\ell_1\ell_2\ell_3\ell_4}(\mathbf{k,k'},\omega=0)]
    a_{\nu_j}^{\ell_2}(\mathbf{k'})a_{\nu_j}^{\ell_3}(\mathbf{-k'}), \label{eq:Gamma_ij}
\end{align}
where $\mathbf{k}\in C_i$ and $\mathbf{k'}\in C_j$ are quasiparticle momenta restricted to the electron or
hole Fermi-surface sheets $C_i$ and $C_j$ and $\nu_i$ and $\nu_j$ are the band indices of these Fermi-surface
sheets. The vertex function in orbital space $\Gamma_{\ell_1\ell_2\ell_3\ell_4}$ describes the particle
scattering of electrons in orbitals $\ell_2,\ell_3$ into $\ell_1,\ell_4$ which is given by RPA in the
fluctuation exchange formalism as
\begin{align}
  &\Gamma_{\ell_1\ell_2\ell_3\ell_4}(\mathbf{k,k'},\omega) = \left[
    \frac{3}{2}\bar{U}^s\chi_1^\text{RPA}(\mathbf{k-k'},\omega)\bar{U}^s+\frac{1}{2}\bar{U}^s \right. \notag \\
  &\quad\quad\quad \left. -\frac{1}{2}\bar{U}^c\chi_0^\text{RPA}(\mathbf{k-k'},\omega)\bar{U}^c+\frac{1}{2}\bar{U}^c
  \right]_{\ell_1\ell_2\ell_3\ell_4}.
\end{align}

The superconducting gap can be factorized into an amplitude $\Delta(T)$ and a normalized symmetry function
$g(\mathbf{k})$. Near $T_c$, the pairing symmetry function $g(\mathbf{k})$ is the stable solution maximizing
the dimensionless pairing strength functional~\cite{Graser09} $\lambda[g(\mathbf{k})]$, which determines
$T_c$. Via the variational method, this is equivalent to solving an eigenvalue problem of the form
\begin{align}
  -\frac{1}{V_G}\sum_j\oint_{C_j} \frac{dS'}{|\mathbf{v}_{\text{F}_j}(\mathbf{k'})|}\Gamma_{ij}(\mathbf{k,k'})g_\mu(\mathbf{k'})
   =\lambda_\mu g_\mu(\mathbf{k}),\label{eq:gapEigenEqa}
\end{align}
where $V_G$ is the volume of 1Fe-BZ, $\mathbf{v}_{\text{F}_j}(\mathbf{k})=\nabla_\mathbf{k}E_j(\mathbf{k})$
is the Fermi velocity on a given Fermi sheet and $dS$ is the area element of the Fermi sheet.   The
eigenfunction $g_\mu(\kk)$ corresponds to the $\mu$th eigenvalue $\lambda_\mu$ and gives the structure of the
gap at the transition. Defining $\mathbf{k}_\perp=(k_\perp,\phi,0)$ in the cylindrical coordinates
$\mathbf{k}=(k_\perp,\phi,k_z)$ and using
$\frac{dS}{|\mathbf{v}_\text{F}(\mathbf{k})|}=\frac{k_\perp^2}{|\mathbf{k}_\perp\cdot\mathbf{v}_\text{F}(\mathbf{k})|}
d\phi dk_z$ is convenient for discretizing the Fermi sheet in parameter form $k_\perp=k_\perp(\phi,k_z)$ into
small patches.~\cite{Crabtree87} A dense ($24\times 12$)-point mesh in parameter space $\{\phi\}\otimes
\{k_z\}$ is used for each Fermi pocket in numerical calculations, implying altogether $n_k\sim 2500$
$\mathbf{k}$ points distributed on all Fermi pockets. After choosing the lattice constant $a$ as the length
unit, eV as the energy unit, and $a\mathrm{eV}/\hbar$ as the velocity unit, the eigenvalue problem
Eq.~(\ref{eq:gapEigenEqa}) reads
\begin{align}
   -\frac{1}{16\pi^3}\sum_j\oint_{C_j} \Gamma_{ij}(\mathbf{k,k'})
     \frac{k_\perp^{'2} d\phi' dk'_z}{|\mathbf{k'}_\perp\cdot\mathbf{v}_\text{F}(\mathbf{k'})|}
     g_\mu(\mathbf{k'})
     =\lambda_\mu g_\mu(\mathbf{k}),\label{eq:gapEigenEqb}
\end{align}
where the normalized eigenfunctions $g_\mu(\mathbf{k})$ are solved numerically by transforming the
integration kernel (for all Fermi sheets $C_i$) into an $n_k\times n_k$ matrix. (The normalization is chosen
so that $\frac{1}{V_G}\sum_j\oint_{C_j}
\frac{dS'}{|\mathbf{v}_{\text{F}_j}(\mathbf{k'})|}[g_\mu(\mathbf{k'})]^2=1$.) Here, we have used the
symmetric pairing vertex $\Gamma_{ij}\equiv
\frac{1}{2}[\Gamma_{ij}(\mathbf{k,k'})+\Gamma_{ij}(\mathbf{k,-k'})]$ for a spin-singlet pairing state since
we want to first examine whether the unconventional superconducting state of the LiFeAs compound and other
Fe-based superconductors is universal and can be explained in the same antiferromagnetic spin-fluctuation
theory before any consideration of triplet pairing or other approaches.

\section{Results for the pairing state}
\subsection{Results for the ARPES-derived Fermi surface}
We now present our solutions to Eq.~(\ref{eq:gapEigenEqb}) for the leading pairing eigenvectors (gap
functions). For this work we fix Hubbard-Hund parameters $U=0.75\ut{eV}$ and $J=0.37U$ and assume
spin-rotational invariance to determine $U'$ and $J'$. These parameters are relatively standard in the
literature making use of the RPA approach to the pairing vertex, and we found that changing them within a
limited range does not change the qualitative aspects of our results for the superconducting state.  The RPA
susceptibility then shows an enhanced incommensurate peak around $\mathbf{q}=\pi(1,0.075,q_z)$
[Fig.~\ref{fig:FSGapsSusc} (c)] or $\mathbf{q}=\pi(1,0.175,q_z)$ [Fig.~\ref{fig:FSGapsSusc} (g)], and the
peak decreases weakly from $q_z=0$ to $q_z=\pi$. (In Fig.~\ref{fig:FSGapsSusc}(g) for filling $n=5.90$,
another peak in the total magnetic susceptibility is at $\mathbf{q}=\pi(0.8,0,q_z)$.) The most important
result as shown in Figs.~\ref{fig:FSGapsSusc}(b) and \ref{fig:FSGapsSusc}(f) and in another representation in
Figs.~\ref{fig:FSGapsSusc}(d) and \ref{fig:FSGapsSusc}(h) is that, using the  ARPES-based band structure for
both fillings considered, we find an $s_\pm$-wave state with anisotropic but full gaps on the electron
(negative gap) and hole (positive gap) pockets.  The other leading eigenvalue corresponds in both cases to a
$d_{xy}$-wave state which is closely competing~\cite{Graser09} but is inconsistent with experiments, such
that we do not investigate it further here.

\begin{figure}
\includegraphics[width=0.9\columnwidth]{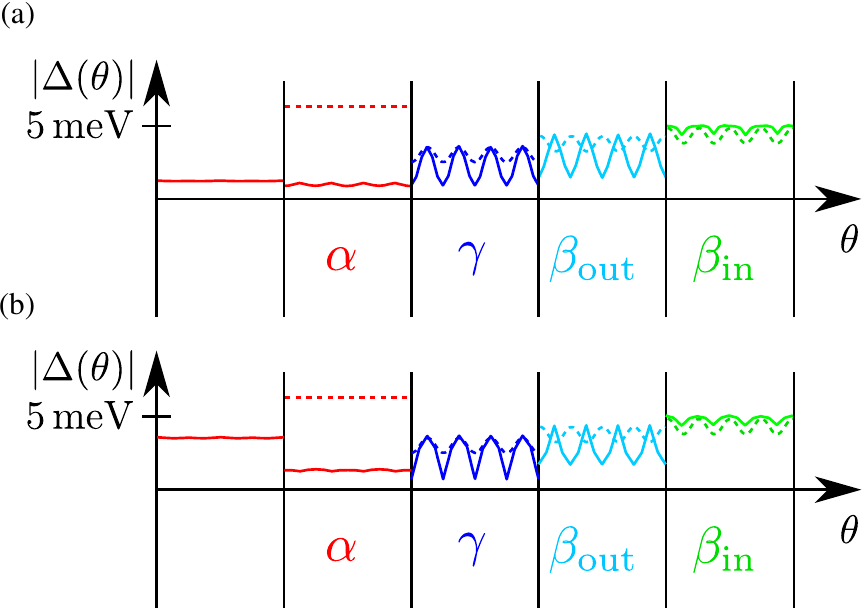}
\caption{(Color online)  Comparison of the gapfunction between the results of the present paper (solid lines)
at $k_z=\pi$ and the experimental findings of $|\Delta(\theta)|$ from Ref. \onlinecite{borisenko_Symmetry}
(dashed lines). Result of the ARPES-derived model at (a) filling $n=6.00$ and (b) at filling $n=5.90$.}
 \label{fig:gapCompare}
\end{figure}

If we now consider the gap functions found on the various pockets in detail, we notice a number of striking
similarities to the experimental results sketched in Fig.~\ref{fig:cartoon}. The full details of the $s_\pm$
gap functions obtained are shown in Figs.~\ref{fig:FSGapsSusc}(b) and  \ref{fig:FSGapsSusc}(f)  and
\ref{fig:FSGapsSusc}(d) and \ref{fig:FSGapsSusc}(h), but for the reader wishing a more concise summary, we
have shown in Fig.~\ref{fig:gapCompare} a schematic comparison of the calculated gaps of the $s_\pm$ states
found at $k_z=\pi$ versus the experimental data, using the angle convention defined in
Fig.~\ref{fig:cartoon}. Taking first the large $\beta$ and $\gamma$ pockets, we see from
Fig.~\ref{fig:gapCompare} that the average gap magnitude is larger on the $\beta$ pockets by a factor of 2 or
so compared to $\gamma$, and the average gap on the inner $\beta$ pocket is about 20\% larger than that on
the outer $\beta$ pocket, as in experiment. The gaps on $\gamma$ and $\beta$ pockets exhibit significant
anisotropy. The minima and maxima on the $\gamma$ pocket are at the same angular positions as in experiment,
and are similar to those from the DFT-based model discussed in the Appendix A. These gap minima are
particularly important as they will determine the momenta of the quasiparticles which dominate
low-temperature measurements {\it if}, as seen in ARPES, the gap on $\gamma$ is the smallest for this system.
Their location along the Fe-Fe bond direction (or the equivalent plane in $\kk$ space) is consistent with
ARPES measurements~\cite{borisenko_Symmetry,Umezawa_2012} as well as with the quasiparticle
interference~\cite{AllenetalScience} and scanning tunneling microscopy experiments~\cite{STMLiFeAs},
according to the interpretation of the latter provided in Ref.~\onlinecite{YanWangvortex}. The gap
oscillations on the $\beta$ pockets are in good agreement with experiment on the $d_{xy}$-orbital-dominated
inner sheets but are apparently $180^\circ$ out of phase with experimental results on the outer electron
sheets. We comment on the origin of this discrepancy below.
\begin{figure*}
  \centering
\begin{pspicture}(0,0)(2\columnwidth,9cm)
 \rput[bl](0.02\columnwidth,4.4){\includegraphics[width=0.96\columnwidth,draft=false]{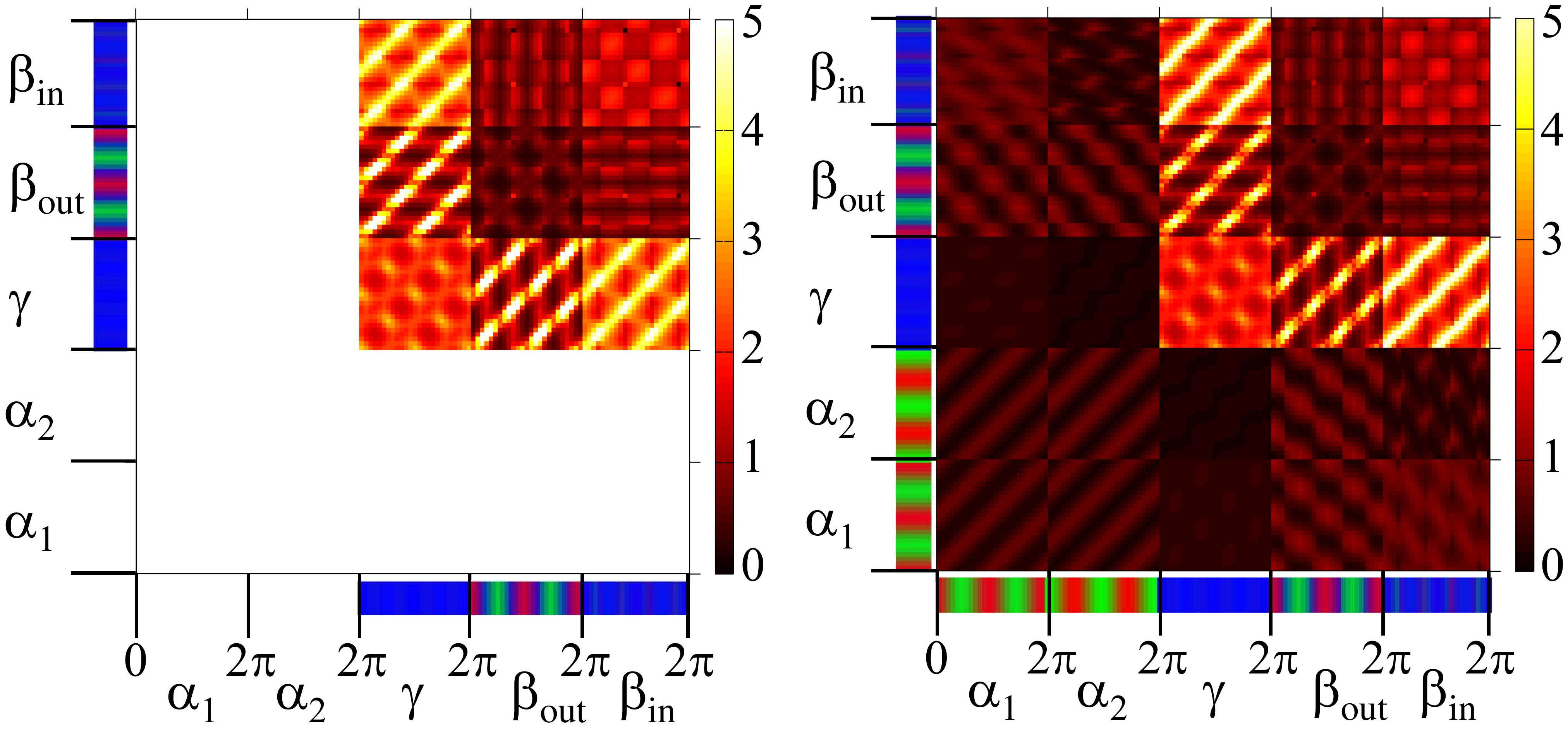}}
 \rput[bl](1.02\columnwidth,4.4){\includegraphics[width=0.96\columnwidth,draft=false]{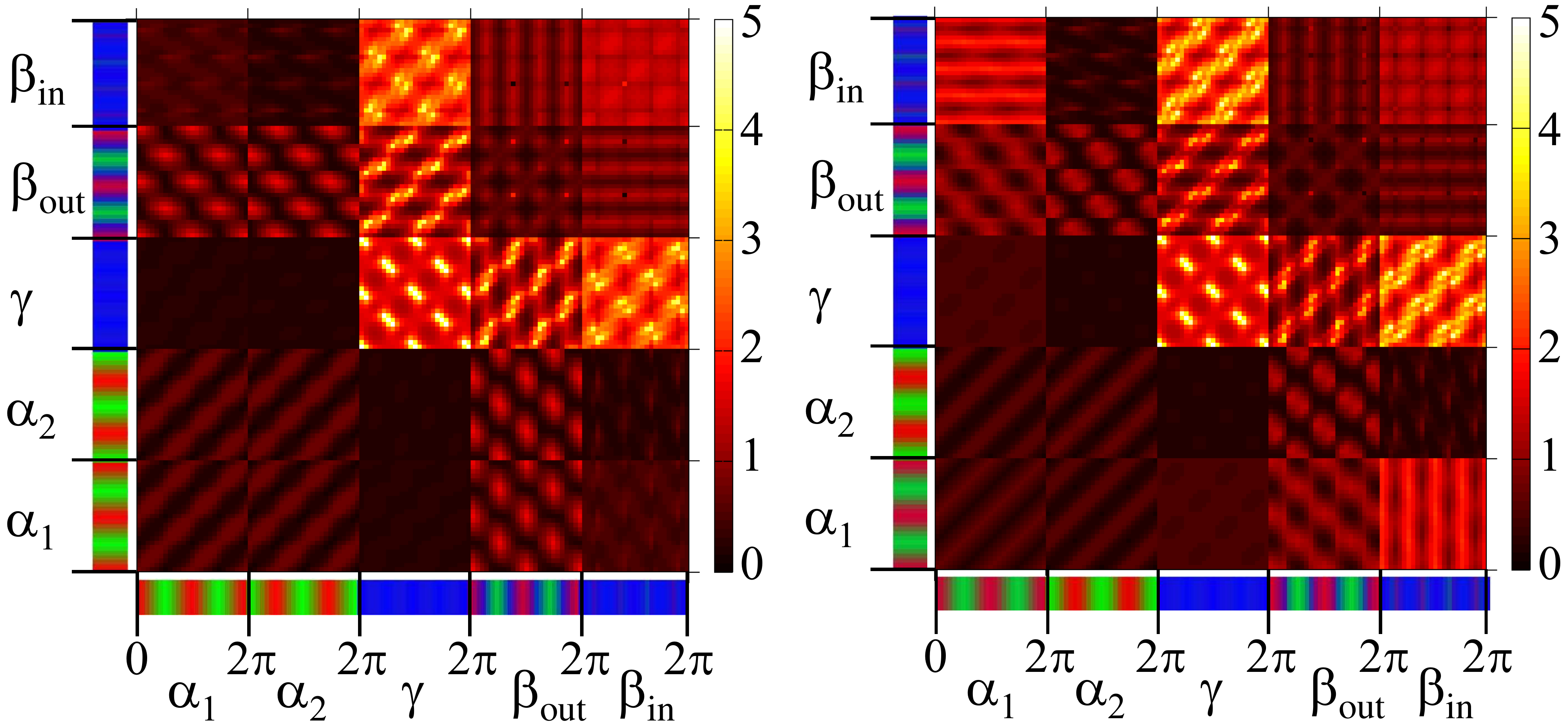}}
 \rput[bl](0.9,8.4){(a) $n=6.00$, $k_z=0$} \rput[bl](5.1,8.4){(b) $n=6.00$, $k_z=\pi$}
 \rput[bl](9.5,8.4){(c) $n=5.90$, $k_z=0$} \rput[bl](13.7,8.4){(d) $n=5.90$, $k_z=\pi$}
 \rput[bl](0.02\columnwidth,0){
    \parbox{0.96\columnwidth}{
    \renewcommand{\arraystretch}{1.4}
    ARPES-based model, filling $n=6.00$
    \begin{tabular*}{0.96\columnwidth}{@{\extracolsep{\fill}}ccc||ccccc}
        \hline \hline
        \multicolumn{1}{c}{band} &\multicolumn{1}{c}{DOS} &\multicolumn{1}{c||}{model gap}
        &\multicolumn{5}{c}{pairing vertex $\Gamma_{\nu\mu}$ } \\
        $\nu$ &$N_\nu(0)$ &$g_\nu$ &$\alpha_{1}$ &$\alpha_{2}$ &$\gamma$ &$\beta_\text{out}$ &$\beta_\text{in}$ \\\hline
        $\alpha_{1}$       &0.01  &0.19        &0.45   &0.44   &0.32   &0.62   &0.76   \\
        $\alpha_{2}$       &0.15  &0.16        &0.44   &0.46   &0.19   &0.60   &0.38   \\
        $\gamma$           &0.70  &0.32        &0.32   &0.19   &2.14   &1.81   &3.11   \\
        $\beta_\text{out}$ &0.34  &$-0.46$     &0.62   &0.60   &1.81   &0.55   &0.77   \\
        $\beta_\text{in}$  &0.15  &$-0.79$     &0.76   &0.38   &3.11   &0.77   &1.38   \\
        \hline \hline
    \end{tabular*}
    }
 }
 \rput[bl](1.02\columnwidth,0){
    \parbox{0.96\columnwidth}{
    \renewcommand{\arraystretch}{1.4}
    ARPES-based model, filling $n=5.90$
    \begin{tabular*}{0.96\columnwidth}{@{\extracolsep{\fill}}ccc||ccccc}
        \hline \hline
        \multicolumn{1}{c}{band} &\multicolumn{1}{c}{DOS} &\multicolumn{1}{c||}{model gap}
        &\multicolumn{5}{c}{pairing vertex $\Gamma_{\nu\mu}$ } \\
        $\nu$ &$N_\nu(0)$ &$g_\nu$ &$\alpha_{1}$ &$\alpha_{2}$ &$\gamma$ &$\beta_\text{out}$ &$\beta_\text{in}$ \\\hline
        $\alpha_{1}$       &0.02  &0.52        &0.43   &0.37   &0.49   &0.82   &1.52   \\
        $\alpha_{2}$       &0.24  &0.20        &0.37   &0.41   &0.20   &0.65   &0.30   \\
        $\gamma$           &0.61  &0.26        &0.49   &0.20   &1.99   &1.29   &2.44   \\
        $\beta_\text{out}$ &0.36  &$-0.37$     &0.82   &0.65   &1.29   &0.50   &0.67   \\
        $\beta_\text{in}$  &0.14  &$-0.69$     &1.52   &0.30   &2.44   &0.67   &1.21   \\
        \hline \hline
    \end{tabular*}
    }
 }
\end{pspicture}
\caption{(Color online) Components of the pairing vertex $\Gamma_{ij}(\mathbf{k,k'})$ matrix resulting in the
pairing function plotted in Fig.~\ref{fig:FSGapsSusc}, from ARPES-based model at (a) and (b) filling $n=6.00$
and (c) and (d) filling $n=5.90$, where the value is proportional to the brightness of the color. The rows
and columns of the tiles of (a)--(d) correspond to Fermi points $\mathbf{k}\in C_i$ and $\mathbf{k'}\in C_j$
where $C_{i,j}$ are the $k_z$ cuts of Fermi sheets $\alpha_{1,2}$, $\gamma$ at $\Gamma$ and
$\beta_\text{out}$, $\beta_\text{in}$ at the $X$ point. Here $k_z=k_z'=0$ for (a) and (c) and $k_z=k_z'=\pi$
for (b) and (d). The angular dependence of the major orbital characters of these Fermi points are labeled by
color as $d_{xz}$ (red), $d_{yz}$ (green), and $d_{xy}$ (blue), as shown in the horizontal and vertical color
bars attached to each panel. The tables show the density of states summed over three dimensions (3D DOS), the
angular averaged pairing vertex
$\Gamma_{\nu\mu}\equiv\sum_{\mathbf{k,k'}}\Gamma(\mathbf{k},\mathbf{k'})/n_\mathbf{k}/n_\mathbf{k'}$ at
$k_z=\pi$ (where $n_\mathbf{k}$ is the number of $\mathbf{k}$ points in the sum), and the approximated model
gap solved from linearized gap equations using the 3D DOS and angular averaged pairing vertex at $k_z=\pi$.}
 \label{fig:GammakkARPESv2}
\end{figure*}
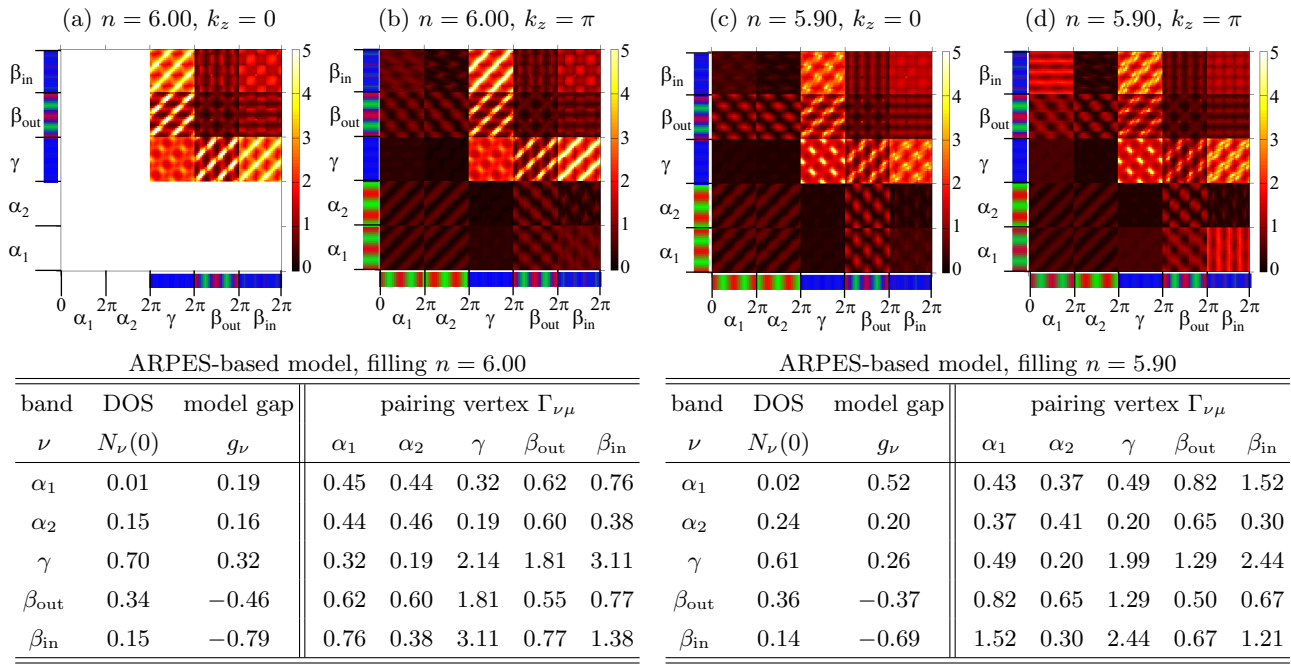

We now discuss the gaps on the $\alpha$  hole pockets.    ARPES\cite{borisenko_Symmetry,borisenko2013} sees
only one band crossing the Fermi level very close to $Z$,   with a large gap of order 6~meV, which we assign
to $\alpha_2$. In our current tight-binding band, which obeys the symmetries of the nonrelativistic DFT
approach, we have always two $\alpha$ pockets or none, as mentioned above.    It may therefore be roughly
appropriate to speak of an average gap on the $\alpha_{1,2}$ pockets in the first analysis.    Within our
calculation with the ARPES-derived band structure, the largest discrepancy with experiment is seen for our
$n=6.00$ calculation, where the $\alpha$ gap is found to be the smallest of all the gaps in the problem. In
the hole-doped case $n=5.90$, the size of the gap on $\alpha_1$ increases significantly, becoming comparable
to experiment, but the gap on $\alpha_2$ remains small.    It is interesting to note that the DFT calculation
(Appendix B), while disagreeing with ARPES on the existence of a $\Gamma$-centered hole pocket, produces a
large gap on both $\alpha_1$ and $\alpha_2$.

\subsection{Analysis of gap sizes in terms of  pairing vertex}
To analyze the origin of the remaining discrepancies with experiment, we investigate the structure of the
effective pair vertex by presenting in Fig.~\ref{fig:GammakkARPESv2} a graphical representation of the
pairing vertex $\Gamma_{ij}(\mathbf{k,k'})$ matrix.  Each block $(i,j)$ in the image represents a matrix
$(\kk,\kk')$ consisting of entries which correspond to the vertex $\Gamma_{ij}(\kk,\kk')$ with  $\kk\in C_i$
and $\kk'\in C_j$. The majority orbital characters along the Fermi surfaces $C_i$  are also indicated in the
Fig.~\ref{fig:GammakkARPESv2}. In the tables below the plots, the densities of states summed over 3D pockets
and scattering vertex components averaged over $(\kk,\kk')$ on the $k_z$ cut are also shown.

For both dopings shown, the brightest set of blocks is that representing scattering processes among the three
largest pockets, $\gamma, \beta_\text{in,out}$.  On average,  it is clear visually in
Fig.~\ref{fig:GammakkARPESv2} and also from the integrated intensities that the dominant scattering processes
within this set of pockets occur for $\gamma$-$\beta_\text{in}$ and $\gamma$-$\gamma$ and, to a lesser
extent, $\gamma$-$\beta_\text{out}$.

There are several interesting conclusions to be drawn from this simple observation. First, one of the crucial
differences between LiFeAs and the ``canonical" 1111 systems which were originally used to deduce general
principles about pairing in the Fe-based superconductors is the existence of a pocket ($\gamma$) with very
large density of states (Fig.~\ref{fig:GammakkARPESv2}) of dominant $d_{xy}$ character. This pocket nests
very poorly with the $\beta$ pockets, as pointed out in Ref.~\onlinecite{borisenko_2010}, but nevertheless
produces the primary pairing interaction leading to superconductivity in part due to the unusually large
$d_{xy}$ content of the $\beta$ pockets in the ARPES-derived bandstructure. This is entirely consistent with
the suggestion that while long-range magnetism is suppressed by the lack of nesting (although this effect
need not rely exclusively on states exactly at the Fermi level), strong magnetic fluctuations remain and are
available for pairing, which is of $s_\pm$ character because the $\gamma$-$\beta$ interactions are repulsive.
It is interesting to note that while the pair scattering processes connecting the $\gamma$ pockets to the
rest of the Fermi surface are large,  the gap on the $\gamma$ pocket itself is not. This is a consequence,
within the theory of multiband superconductivity, of the large density of states on the $\gamma$ pocket, as
discussed in Sec.~\ref{subsec:toymodel}.

Second, we note that the intraband scattering $\gamma$-$\gamma$ is also quite strong.  These are small-$\q$
processes which may be responsible for the tendency to ferromagnetism seen in these
systems~\cite{bBuchner2013}. Although we do not see enhancement of the {\it total} magnetic susceptibility
near $\mathbf{q}=0$ (Fig.~\ref{fig:FSGapsSusc}), there are evidently intraorbital susceptibilities including
$\chi_{xy,xy,xy,xy}$ which are large at small $\mathbf{q}$, and the partial density of states $N_{\gamma}(0)$
is the largest among all pockets.

Finally, we note that  the  strong angular dependence of the vertex is induced by the variation of the
orbital content in general, and the $d_{xz}/d_{yz}$ content in particular.   As discussed in
Refs.~\onlinecite{Maier09} and~\onlinecite{Kemper10},  there is a strong tendency for pair scattering between
like orbitals to be enhanced, other effects being equal, accounting for the large $\gamma$-$\beta_\text{in}$
scattering. But even in this case subdominant $xz/yz$ orbitals are present on the $\beta_\text{in}$ sheets
which lead to the observed modulation via the matrix elements which occur in Eq.~(\ref{eq:Gamma_ij}).

To understand the angular oscillations within a phenomenological picture, Maiti
\textit{et~al.}~\cite{Maiti12} fitted the gaps on the electron pockets measured by Umezawa \textit{et~al.}
with the angle dependence
\begin{subequations}
\begin{align}
  \Delta_\text{inner}(\theta) &=\Delta_0(1+r_2|\cos2\theta|+r_4\cos4\theta), \\
  \Delta_\text{outer}(\theta) &=\Delta_0(1-r_2|\cos2\theta|+r_4\cos4\theta),
\end{align}
\end{subequations}
where $\theta$ is defined in the caption of Fig.~\ref{fig:cartoon} (measured from dashed-line direction), and
they found (i) $r_2>0$ and (ii) $r_4>\frac{1}{4}r_2$. Point (i) is equivalent to
$\Delta_\text{inner}>\Delta_\text{outer}$, which is measured by both ARPES experiments, and our results from
both tight-binding models also agree with point (i). Point (ii) is related to the in-phase feature and the
orientation of gap maxima on both pockets because, first, at $\theta=0$, $d\Delta_\text{inner}/d\theta=0$ and
$d\Delta_\text{outer}/d\theta=0$ and, second, $d^2\Delta_\text{inner}/d\theta^2=-4(r_2+4r_4)$ and
$d^2\Delta_\text{outer}/d\theta^2=4(r_2-4r_4)$. $r_4>\frac{1}{4}r_2$ means both inner and outer pockets have
maxima at $\theta=0$, and hence they are in phase. Considering $r_2>0$ and the gap on the outer pocket
oscillates stronger than the inner pocket (larger curvature at $\theta=0$) in our results, a reasonable range
for $r_4$ at all $k_z$ is  $r_4\sim -\frac{1}{4}r_2$. The sign of $r_2$ is determined by the angle dependence
of the pairing interaction and is positive in the case where the interaction between electron and hole
pockets is dominant.~\cite{Maiti12} Our numerical results suggest the same conclusion as the ARPES
experiments.  The discrepancy in the phase of the oscillations on the outer $\beta$ pocket is attributable to
the ``wrong" sign of the more sensitive parameter~\cite{Maiti12} $r_4$ obtained within our calculations.

We now turn to the more delicate issue of the pairing-vertex components connecting the $\alpha$ pockets to
the rest of the Fermi surface.   It is clear from both the plots and table corresponding to
Figs.~\ref{fig:GammakkARPESv2}(a) and  \ref{fig:GammakkARPESv2}(b) that these are negligible in the
compensated case $n=6.00$, in Fig.~\ref{fig:GammakkARPESv2}(a) simply because $H_0^\text{ARPES}$ contains no
$\alpha$ pockets at $k_z=0$, and in Fig.~\ref{fig:GammakkARPESv2}(b) because the densities of states on these
closed 3D pockets are extremely small. In 2D models, where densities of states tend to be weakly dependent on
pocket size, these effects are suppressed.    We discuss the connection of the small gap on the $\alpha$
pockets to the corresponding components of the vertex below.  For the moment,  we note simply that the effect
of hole doping to $n=5.90$ shown in Figs.~\ref{fig:GammakkARPESv2}(c) and \ref{fig:GammakkARPESv2}(d) clearly
enhances the scattering of pairs on the $\alpha$ pockets to the $\beta$ pockets, particularly to
$\beta_\text{in}$. As seen in Figs.~\ref{fig:FSGapsSusc}(a) and \ref{fig:FSGapsSusc}(e),  hole doping by a
small amount (5\% Fe) transforms the small $Z$-centered $\alpha$ pockets into two narrow concentric tubes and
thereby enhances the DOS on the $\alpha$ pockets.  While the $n=5.90$ case  is nominally inconsistent with
the ARPES observation of no $\alpha$-type Fermi surfaces at $k_z=0$, it is clear that the determination of
the hole dispersion near $\Gamma$-$Z$ becomes quite challenging when the bands are grazing the Fermi level.
It is significant that the results for the DFT analysis given in Appendix B also give large gaps on the
$\alpha$ pockets, although the Fermi surface of the hole pockets disagrees qualitatively with ARPES. Taken
together, these results suggest that the $\alpha$-$\beta$ interaction is enhanced and the gap on the $\alpha$
pocket is large only  if states near $\Gamma$ of $xz/yz$ character contribute strongly to pairing. This
occurs when the Fermi surface includes an open (cylindrical) $\alpha_1$ pocket and also when the range of
pairing is expanded to include states away from the Fermi level, as discussed below.

\subsection{Discussion: Toy model for gap sizes} \label{subsec:toymodel}
To understand the relative sizes of the gaps on the various Fermi surface sheets, one needs to combine
knowledge of the pairing vertex function discussed above with the densities of states on each sheet.  Here, a
simplified picture can help us understand why certain gaps are large and others are small.  We neglect for
this discussion the momentum dependence of the gap eigenfunctions, densities of states, and vertices over the
individual Fermi surface sheets.  If we are primarily interested in gap {\it sizes}, a good approximation to
the gap equation~(\ref{eq:gapEigenEqb}) is then given by
\begin{eqnarray}\label{isotropic_gapeqn}
 \lambda g_i &\approx& \sum_j -g_j N_j \Gamma_{ij}  ,
\end{eqnarray}
where $g_i$ now denotes the gap on the $i$th band and so on.   We begin by discussing the question of how the
gap on tiny $Z$-centered (or $\Gamma$-centered) hole pockets can become large, as seen by
ARPES\cite{borisenko_Symmetry}. Were intraband scattering processes dominant, the tiny DOS on the $\alpha$
pockets would generically create an extremely small gap.  Since interband scattering is more important, in
the situation where the DOS on the $\alpha$ pockets is small, the gap on $\alpha$ will be determined by
scattering from the other major bands, in particular $\gamma$ and $\beta_\text{in,out}$ as seen in
Fig.~\ref{fig:GammakkARPESv2}.

In such a situation, we have approximately
\begin{equation}\label{toymodel}
    \lambda g_\alpha \approx- g_\gamma N_\gamma \Gamma_{\alpha\gamma} - \sum_\nu g_{\beta_\nu} N_{\beta_\nu} \Gamma_{\alpha\beta_\nu} ,
\end{equation}
where $\nu$ sums over both inner and outer $\beta$ pockets.  Since the state is an $s_\pm$ state driven by
repulsive interband interactions, the first and second terms tend to cancel each other.  Large gaps can then
be obtained if parameters are chosen such that the contribution from the $\gamma$ pocket is minimized. As we
have seen above, however, in the current ARPES-derived model, while the scattering of $\alpha$ states to the
$\beta$ pockets is much stronger, the $\gamma$ density of states is significantly larger, such that the two
terms in Eq.~(\ref{toymodel}) are comparable and therefore mostly cancel each other. As can be seen by
comparing the hole-doped case with the compensated case with the tables for $\Gamma_{\mu\nu}$ in
Fig.~\ref{fig:GammakkARPESv2}, the main effect of the hole doping on the balance of the two terms in
Eq.~(\ref{toymodel}) is due to the enhancement of $\Gamma_{\alpha,\beta_\text{in}}$ by a factor of 2.

\section{Conclusions}
We have performed 3D calculations of the superconducting pair state in the LiFeAs compound, one of the few
materials where ARPES experiments  indicate significant gap anisotropy, possibly due to reduced diffuse
scattering from the very clean, nonpolar surface. Since the inner hole pockets of the  Fermi surface of this
material are thought to be strongly renormalized by interactions, we used as the input a tight-binding model
fit to ARPES data reproducing both the band structure and orbital polarization measurements at the Fermi
surface. Our calculations find a gap structure which changes sign between the hole and electron pockets and
reproduce semiquantitatively the relative gap sizes on the three largest pockets,   along with the
oscillatory behavior seen. We performed a careful analysis of the structure of the pair-scattering vertex to
understand the structure of these pair states. The gap function observed by ARPES on the main pockets can
then be understood entirely in terms of the repulsive interband interactions within the spin-fluctuation
approach. On the outer electron pocket, a difference in the sign of the oscillations with respect to
experiment can be traced to a term in the phenomenology of Maiti \textit{et al.}~\cite{Maiti12} which depends
very sensitively on the balance between intra- and interpocket interactions.

Our results differ from experiment in one important respect, namely the small size of the gap on the inner
hole ($\alpha$) pockets  we find, in contrast to the large gap observed in
Ref.~\onlinecite{borisenko_Symmetry}.  We discussed here, and in Appendices A and B, various model Fermi
surfaces which tend towards giving significantly larger $\alpha$ gaps and deduced that inclusion of the
$xz/yz$ states in the pairing near the $\Gamma$ point of the Brillouin zone appears to be essential.  While
these models do not appear to be fully consistent with the Fermi surface found by ARPES,  they point the way
towards identifying missing ingredients in the theory.  In particular, since the $\alpha$ pockets in this
material are tiny and very close to a Lifschitz transition, it may be necessary in this system to account for
states slightly away from the Fermi level in order to reproduce the overall gap structure.

Due to the remarkable agreement of the robust part of the  gap structure on the main pockets,  we conclude
that the pairing in LiFeAs has essentially the same origin as in other Fe-based superconductors, despite the
fact that there is no nesting apparent at the Fermi surface.   We point out that the main difference between
LiFeAs and the paradigmatic 1111 systems is the predominance of the scattering between the hole $\gamma$
Fermi pocket and the electron $\beta$ pockets, all of which have substantial $xy$ orbital character; pure
$xz/yz$ scattering is subdominant. A strong $d_{xy}$ intrapocket interaction may explain the ferromagnetic
correlations observed in experiment, despite the lack of a $\mathbf{q}=0$ peak in the total magnetic
susceptibility.

\begin{acknowledgments}
The authors gratefully acknowledge useful discussions with B. M. Andersen,  T. Berlijn, D. A. Bonn, A.
Chubukov, A. Coldea, A. Damascelli, J. C. Davis, I. Eremin, M. N. Gastiasoro, J. A. Hoffman, H. Jeschke, S.
Johnston, M. Khodas,  M. Korshunov,  W. Ku, G. Levy, I. I. Mazin,  M. Tomi\'{c}, R. Valent\'{i}, and
M.Watson. P.J.H., Y.W., and A.K. were supported by Grant No. DOE DE-FG02-05ER46236. V.B.Z., S.V.B. and B.B.
acknowledge support under Grants No.~ZA 654/1-1, No. BO1912/2-2, and No. BE1749/13. A portion of this
research was conducted at the Center for Nanophase Materials Sciences, which is sponsored at Oak Ridge
National Laboratory by the Scientific User Facilities Division, Office of Basic Energy Sciences, U.S.
Department of Energy.
\end{acknowledgments}

\appendix
\section{Electronic structure of LiFeAs from Density Functional Theory} \label{appendix:DFTbands}
\begin{figure}
 \centering
 \includegraphics[width=0.88\columnwidth]{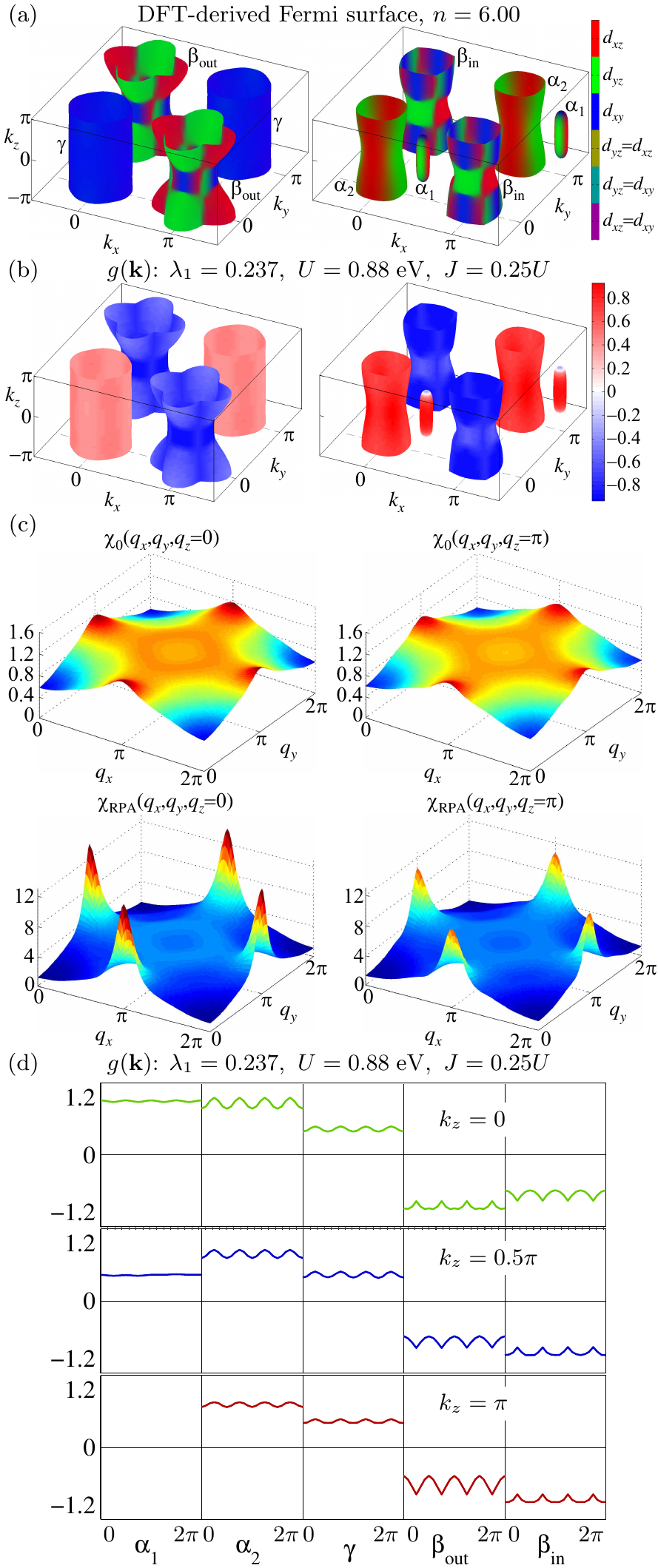}
\caption{(Color online) (a)  Fermi surface of LiFeAs from ten-orbital $H_0^\text{DFT}$ plotted in the
coordinates of the one-Fe Brillouin zone as two sets, outer (left) and inner (right) pockets. Majority
orbital weights are labeled by colors as shown. Note the small innermost, hole pocket $\alpha_1$ with the
rotation axis $\Gamma$-$Z$ (or $M$-$A$) has been artificially displaced from its position along the $k_x$
axis for better viewing.  (b) The gap symmetry functions $g(\mathbf{k})$ corresponding to the leading
eigenvalues ($s_\pm$ wave) with interaction parameters shown in the figure. (c) The corresponding
noninteracting spin susceptibility [$\chi_s(\mathbf{q},\omega=0)$ with $U=0,J=0$] and RPA spin susceptibility
[$\chi_s(\mathbf{q},\omega=0)$ with the same $U,J$ as in (b) and (f)] at $q_z=0,\pi$. (d) The angle
dependence of $g(\mathbf{k})$ on the pockets indicated at $k_z=0,0.5\pi,\pi$.}
 \label{fig:FSGapsSuscDFT}
\end{figure}

The band structure from DFT for the LiFeAs parent compound is calculated using the quantum \textsc{espresso}
package. The experimentally determined lattice parameters used in the calculation are taken from Table I in
Ref.~\onlinecite{Tapp08}, including lattice constants $a=3.7914\ut{\AA}$, $c=6.3639\ut{\AA}$ and the internal
coordinates for the Li atoms $z_\text{Li}=0.8459$ and the As atoms $z_\text{As}=0.2635$. Next we obtain the
DFT derived ten-orbital tight-binding Hamiltonian model $H_0^\text{DFT}$ by projecting the bands near the
Fermi energy on the ten $3d$-orbitals of the two Fe atoms in the primitive cell of the LiFeAs crystal using
maximally localized Wannier functions computed using the \textsc{wannier90} package. The Fermi surface from
this model is shown in Fig.~\ref{fig:FSGapsSuscDFT}(a), where the colors encode the orbital character. The
Fermi surface sheets of the ten-orbital model are plotted using a \emph{repeated-zone} scheme of the two-Fe
Brillouin zone (2Fe-BZ) in the coordinates $(k_x,k_y,k_z)$ of the 1Fe-BZ, and the cube in $\mathbf{k}$ space
in Fig.~\ref{fig:FSGapsSuscDFT}(a) encloses the volume of the 1Fe-BZ. This representation is convenient for
later calculation since the susceptibility is only a periodic function in the 1Fe-BZ. For the convenience of
later discussion, we denote the two hole pockets at the $\Gamma(0,0,0)$ [or $M(\pi,\pi,0)$] point as
$\alpha_1$/$\alpha_2$ and two electron pockets at the $X$ (or $Y$) point as
$\beta_\text{out}$/$\beta_\text{in}$. The DOS at the Fermi level is shown in Table~\ref{tab:dosDFTandARPES},
in comparison with that of ARPES-derived model.

\section{Pairing state for DFT-derived Fermi surface} \label{appendix:DFTpairing}
Although the Fermi surface predicted by DFT differs in some essential respects from that found in ARPES, it
is nevertheless useful to calculate the gap functions which arise within the spin-fluctuation theory for this
electronic structure to get a sense of how much the gap varies for small changes in the electronic structure
and to compare with earlier 2D theoretical calculations using a DFT-derived Fermi surface\cite{Platt11}. As
shown in Fig.~\ref{fig:FSGapsSuscDFT}, for $U=0.88\ut{eV}$ and $J=0.25U$,  we find an $s_\pm$-wave state
($\lambda_1=0.237$) with anisotropic full gaps on the electron (negative gap) and hole (positive gap)
pockets, as shown in Figs.~\ref{fig:FSGapsSuscDFT}(b) and \ref{fig:FSGapsSuscDFT}(d). (The second eigenvalue
$\lambda_2=0.1006$ is a $d_{x^2-y^2}$-wave state.) The $s_\pm$-wave state is driven by the enhanced {\it
commensurate} peak at $\mathbf{q}=(\pi,0,q_z)$ in the RPA susceptibility, see
Fig.~\ref{fig:FSGapsSuscDFT}(c). This peak value has a moderate $q_z$ dependence and becomes smaller at
$q_z=\pi$, which means the gap structure will not change too much along $k_z$.

The gaps on the hole pockets $\alpha_2$ and $\gamma$ exhibit moderate $k_z$ dependence. The gap minima on the
$\gamma$ pocket are in the $k_x,k_y$ or Fe-Fe directions. The gap on the closed $\alpha_1$ pocket is among
the largest ones, although the DOS of the $\alpha_1$ pocket is the smallest, and this gap shows strong $k_z$
dependence near the pole of the pocket. Next, the gaps on the inner $\beta_\text{in}$ and outer
$\beta_\text{out}$ pockets seem to be intertwined and correlated: near $k_z=0$, the gap on the inner pocket
is smaller than the gap on the outer pocket, while near $k_z=\pi$ the order is flipped [see
Fig.~\ref{fig:FSGapsSuscDFT}(d) for gaps at $k_z=0,\pi$], but they coincide at the points where two Fermi
pockets touch each other. Last, while the gap magnitudes from our full 3D calculation are substantially
similar to those obtained using 2D functional renormalization group calculations by  Platt \textit{et
al.}~\cite{Platt11} at $k_z=0$, at $k_z=\pi$ we find qualitatively different hole pocket gaps, indicating the
importance of 3D pair-scattering processes.

\begin{table}
\caption{LiFeAs density of states (two Fe atoms, two spins) at the Fermi level from (a) the ten-orbital
DFT-based model $H_0^\text{DFT}$ and (b) the ARPES-based model $H_0^\text{ARPES}$ along with (c) the density
of states in (b) scaled by a factor $r=0.5$.}
 \label{tab:dosDFTandARPES}
\renewcommand{\arraystretch}{1.4}
\begin{tabular*}{\columnwidth}{@{\extracolsep{\fill}}l|cccccc}
\hline \hline
    &$\alpha_{1}$  &$\alpha_{2}$  &$\gamma$  &$\beta_\text{out}$  &$\beta_\text{in}$   &total \\\hline
(a) &0.040         &0.554         &0.660     &0.610          &0.377          &2.241  \\
(b) &0.038         &0.592         &2.782     &1.377          &0.594          &5.383  \\
(c) &0.019         &0.296         &1.391     &0.689          &0.297          &2.692  \\
\hline \hline
\end{tabular*}
\end{table}

\section{Fitting parameters for the ten-orbital tight-binding model $H_0^\text{ARPES}$} \label{appendix:TBmodel}
In the following, we give the Hamiltonian matrix of the tight-binding model $H_0^\text{ARPES}$ from
Ref.~\onlinecite{Eschrig09} (with corrections and minor changes) and the hopping parameters
$t_{\ell\ell'}^{rst}$ which are obtained by fitting the ARPES measured band structure for LiFeAs using that
tight-binding model. The hopping parameters were tuned to optimally reproduce a vast set of experimental data
measured along high symmetry directions as well as complete Fermi surface maps, cutting the band structure at
arbitrary angles to crystallographic axes. One such map is shown in Fig.~\ref{fig:fit}. To pin down $k_z$
dispersions, high symmetry cuts measured with different excitation energies were used. Here $\ell,\ell'$ are
orbital indices with $1=d_{xy}$, $2=d_{x^2-y^2}$, $3=id_{xz}$, $4=id_{yz}$, $5=d_{z^2}$ for the first Fe
within the unit cell and $6=d_{xy}$, $7=d_{x^2-y^2}$, $8=-id_{xz}$, $9=-id_{yz}$, $10=d_{z^2}$ for the second
Fe. $r,s,t$ are integers denoting a hopping distance $r\mathbf{T}_x+s\mathbf{T}_y+t\mathbf{R}_3$ where
$\mathbf{R}_1$, $\mathbf{R}_2$, $\mathbf{R}_3$ are lattice basis vectors and $\mathbf{T}_x$, $\mathbf{T}_y$
are basis vectors for the one-Fe unit cell. Specifically, we have
$\mathbf{T}_x=\frac{1}{2}(\mathbf{R}_1-\mathbf{R}_2)$, $\mathbf{T}_y=\frac{1}{2}(\mathbf{R}_1+\mathbf{R}_2)$,
and accordingly in the reciprocal space, we have $k_1=k_x+k_y$, $k_2=-k_x+k_y$, $k_3=k_z$, where the
wave-number components are scaled by choosing the lattice constant $a=1$. The entire calculation is done with
$k_{1,2,3}$ (in 2Fe-BZ) and then plotted with $k_{x,y,z}$ (in 1Fe-BZ using a repeated-zone scheme), such as,
for example, in Fig.~\ref{fig:FSGapsSusc}(a). $H_0^\text{ARPES}$ is given in the block matrix form as
follows:
\begin{align} \label{eq:H0tbhARPES}
    H_0^\text{ARPES}&=
  \begin{pmatrix}
    H^{++}  &H^{+-}\\
    H^{+-*} &H^{++*}
  \end{pmatrix}.
\end{align}
Here an asterisk ($^*$) means complex conjugate. Each element of $H^{++},H^{+-}$ is given in two parts: the
2D part and the 3D part.

For the 2D part of the Hamiltonian,
\begin{align}
  H^{++}_{11} &= \epsilon_1 + 2t^{11}_{11}(\cos k_1 + \cos k_2) + 2t^{20}_{11}(\cos 2k_x + \cos 2k_y)
   ,\notag\\
  H^{++}_{12} &= 0
   ,\notag\\
  H^{++}_{13} &= 2it^{11}_{13}(\sin k_1 - \sin k_2)
   ,\notag\\
  H^{++}_{14} &= 2it^{11}_{13}(\sin k_1 + \sin k_2)
   ,\notag\\
  H^{++}_{15} &= 2t^{11}_{15}(\cos k_1 - \cos k_2)
   ,\notag\\
  H^{++}_{22} &= \epsilon_2 + 2t^{11}_{22}(\cos k_1 + \cos k_2)
   ,\notag\\
  H^{++}_{23} &= 2it^{11}_{23}(\sin k_1 + \sin k_2)
   ,\notag\\
  H^{++}_{24} &= 2it^{11}_{23}(-\sin k_1 + \sin k_2)
   ,\notag\\
  H^{++}_{25} &= 0
   ,\notag\\
  H^{++}_{33} &= \epsilon_3 + 2t^{11}_{33}(\cos k_1 + \cos k_2) + 2t^{20}_{33}\cos 2k_x  \notag\\
   & + 2t^{02}_{33}\cos 2k_y
   +4t^{22}_{33}\cos 2k_x \cos 2k_y
   ,\notag\\
  H^{++}_{34} &= 2t^{11}_{34}(\cos k_1 - \cos k_2)
   ,\notag\\
  H^{++}_{35} &= 2it^{11}_{35}(\sin k_1 + \sin k_2)
   ,\notag\\
  H^{++}_{44} &= \epsilon_3 + 2t^{11}_{33}(\cos k_1 + \cos k_2) + 2t^{02}_{33}\cos 2k_x \notag\\
   & + 2t^{20}_{33}\cos 2k_y + 4t^{22}_{33}\cos 2k_x \cos 2k_y
   ,\notag\\
  H^{++}_{45} &= 2it^{11}_{35}(\sin k_1 - \sin k_2)
   ,\notag\\
  H^{++}_{55} &= \epsilon_5
   ,\notag\\
  H^{++}_{ji} &= (H^{++}_{ij})^*
   .
\end{align}
\begin{align}
  H^{+-}_{16} &= 2t^{10}_{16}(\cos k_x + \cos k_y) \notag\\
   & + 2t^{21}_{16}[(\cos k_1 + \cos k_2)(\cos k_x + \cos k_y) \notag\\
   & - \sin k_1(\sin k_x + \sin k_y) + \sin k_2(\sin k_x - \sin k_y)]
   ,\notag\\
  H^{+-}_{17} &= 0
   ,\notag\\
  H^{+-}_{18} &= 2it^{10}_{18} \sin k_x
   ,\notag\\
  H^{+-}_{19} &= 2it^{10}_{18} \sin k_y
   ,\notag\\
  H^{+-}_{1,10} &= 0
   ,\notag\\
  H^{+-}_{27} &= 2t^{10}_{27} (\cos k_x + \cos k_y)
   ,\notag\\
  H^{+-}_{28} &= -2it^{10}_{29} \sin k_y
   ,\notag\\
  H^{+-}_{29} &= 2it^{10}_{29} \sin k_x
   ,\notag\\
  H^{+-}_{2,10} &= 2t^{10}_{2,10} (\cos k_x - \cos k_y)
   ,\notag
\end{align}
\begin{align}
  H^{+-}_{38} &= 2t^{10}_{38}\cos k_x + 2t^{10}_{49}\cos k_y \notag\\
   &\mspace{-30mu} + 2t^{21}_{38}[(\cos k_1 + \cos k_2)\cos k_x - (\sin k_1 - \sin k_2)\sin k_x] \notag\\
   &\mspace{-30mu} + 2t^{21}_{49}[(\cos k_1 + \cos k_2)\cos k_y - (\sin k_1 + \sin k_2)\sin k_y]
   ,\notag\\
  H^{+-}_{39} &= 0
   ,\notag\\
  H^{+-}_{3,10} &= 2it^{10}_{4,10} \sin k_y
   ,\notag\\
  H^{+-}_{49} &= 2t^{10}_{49}\cos k_x + 2t^{10}_{38}\cos k_y \notag\\
   &\mspace{-30mu} + 2t^{21}_{49}[(\cos k_1 + \cos k_2)\cos k_x - (\sin k_1 - \sin k_2)\sin k_x] \notag\\
   &\mspace{-30mu} + 2t^{21}_{38}[(\cos k_1 + \cos k_2)\cos k_y - (\sin k_1 + \sin k_2)\sin k_y]
   ,\notag\\
  H^{+-}_{4,10} &= 2it^{10}_{4,10} \sin k_x
   ,\notag\\
  H^{+-}_{5,10} &= 0
   .
\end{align}

For the 3D part of the Hamiltonian,
\begin{align}
  H^{++}_{11} &= H^{++}_{11} + [2t^{001}_{11}+4t^{111}_{11}(\cos k_1 + \cos k_2) \notag\\
   & + 4t^{201}_{11}(\cos 2k_x + \cos 2k_y)]\cos k_z
   ,\notag\\
  H^{++}_{13} &= H^{++}_{13} - 4t^{201}_{14}\sin 2k_y \sin k_z
   ,\notag\\
  H^{++}_{14} &= H^{++}_{14} - 4t^{201}_{14}\sin 2k_x \sin k_z
   ,\notag\\
  H^{++}_{33} &= H^{++}_{33} + [2t^{001}_{33}+4t^{201}_{33}\cos 2k_x + 4t^{021}_{33}\cos 2k_y]\cos k_z
   ,\notag\\
  H^{++}_{44} &= H^{++}_{44} + [2t^{001}_{33}+4t^{021}_{33}\cos 2k_x + 4t^{201}_{33}\cos 2k_y]\cos k_z
   ,\notag\\
  H^{+-}_{16} &= H^{+-}_{16} + 4t^{101}_{16}(\cos k_x + \cos k_y)\cos k_z \notag\\
   & + 2t^{121}_{16}\{[\cos (k_1+k_y) + \cos (k_1+k_x)]\exp(ik_z) \notag\\
   & + [\cos (k_2+k_y) + \cos (k_2-k_x)]\exp(-ik_z)\}
   ,\notag\\
  H^{+-}_{18} &= H^{+-}_{18} + 4it^{101}_{18}\sin k_x \cos k_z - 4t^{101}_{19}\sin k_y \sin k_z \notag\\
   &\mspace{-30mu} + 2it^{211}_{19}[\sin (k_1+k_y) \exp(ik_z) - \sin (k_2+k_y) \exp(-ik_z)]
   ,\notag\\
  H^{+-}_{19} &= H^{+-}_{19} + 4it^{101}_{18}\sin k_y \cos k_z - 4t^{101}_{19}\sin k_x \sin k_z \notag\\
   &\mspace{-30mu} + 2it^{211}_{19}[\sin (k_1+k_x) \exp(ik_z) + \sin (k_2-k_x) \exp(-ik_z)]
   ,\notag\\
  H^{+-}_{38} &= H^{+-}_{38} + 4(t^{101}_{38}\cos k_x + t^{101}_{49}\cos k_y)\cos k_z \notag\\
   &\mspace{-30mu} + 2t^{121}_{38}[\cos (k_1+k_x) \exp(ik_z) + \cos (k_2-k_x) \exp(-ik_z)] \notag\\
   &\mspace{-30mu} + 2t^{121}_{49}[\cos (k_1+k_y) \exp(ik_z) + \cos (k_2+k_y) \exp(-ik_z)]
   ,\notag
\end{align}
\begin{align}
  H^{+-}_{39} &= H^{+-}_{39} + 4it^{101}_{39}(\cos k_x + \cos k_y)\sin k_z
   ,\notag\\
  H^{+-}_{49} &= H^{+-}_{49} + 4(t^{101}_{49}\cos k_x + t^{101}_{38}\cos k_y)\cos k_z \notag\\
   &\mspace{-30mu}  + 2t^{121}_{49}[\cos (k_1+k_x) \exp(ik_z) + \cos (k_2-k_x) \exp(-ik_z)] \notag\\
   &\mspace{-30mu}  + 2t^{121}_{38}[\cos (k_1+k_y) \exp(ik_z) + \cos (k_2+k_y) \exp(-ik_z)]
   .
\end{align}

The numerical values for hopping parameters in units of eV are as follows. For the 2D part,
\begin{align*}
    \epsilon_1  &=  0.020,     \ \epsilon_2 = -0.2605,     \ \epsilon_3 = -0.0075,\\     \ \epsilon_5 &= -0.3045,
    t^{11}_{11} =  0.030,    \ t^{10}_{16} = -0.0185,\\    \ t^{20}_{11} &= -0.010,    \ t^{21}_{16} =  0.0035,
    t^{11}_{13} = -0.0635i,\\  \ t^{10}_{18} &=  0.155i,   \ t^{11}_{15} = -0.090,    \ t^{10}_{27} = -0.2225,\\
    t^{11}_{22} &=  0.070,    \ t^{10}_{29} = -0.1925i,   \ t^{11}_{23} = -0.010i,\\   \ t^{10}_{2,10} &=  0.1615,
    t^{11}_{33} =  0.152,    \ t^{10}_{38} =  0.050,\\    \ t^{20}_{33} &= -0.004,    \ t^{21}_{38} =  0.040,
    t^{02}_{33} = -0.051,\\    \ t^{10}_{49} &=  0.210,    \ t^{22}_{33} = -0.005,    \ t^{21}_{49} = -0.053,\\
    t^{11}_{34} &=  0.090,    \ t^{10}_{4,10}= 0.0995i,   \ t^{11}_{35} =  0.1005i.
\end{align*}
For the 3D part,
\begin{align*}
    t^{101}_{16} &= -0.004, \ t^{001}_{11} = 0.0105, \ t^{111}_{11} = 0,\\    \ t^{201}_{11} &= 0, \ t^{201}_{14} = 0,
    t^{001}_{33} = -0.003,\\ \ t^{201}_{33} &= 0,     \ t^{021}_{33} = 0.0105,\ t^{121}_{16} = 0,\\ \ t^{101}_{18} &= 0,
    t^{101}_{19} = 0,      \ t^{211}_{19} = 0,\\     \ t^{101}_{38} &= 0.0115,\ t^{121}_{38} = 0, \ t^{101}_{39} = 0,\\
    t^{101}_{49} &= 0,      \ t^{121}_{49} = 0.
\end{align*}

Some hopping parameters $t^{rst}_{\ell\ell'}$ are purely imaginary numbers because the $d_{xz}$ and $d_{yz}$
orbitals are multiplied by the imaginary unit factor to get the real Hamiltonian matrix. However, if one were
interested in orbital resolved susceptibility or pairing vertex function, real orbitals are more
meaningful,~\cite{Kemper10} so we can introduce a gauge transformation to transform to real orbitals by the
matrix $S={\rm diag}(1,1,i,i,1,1,1,-i,-i,1)$, and the transformed Hamiltonian is
$\tilde{H}_0^\text{ARPES}=S^{-1}H_0^\text{ARPES}S$.

%

\end{document}